\documentclass[article,balancelastpage,twocolumn,prl]{revtex4}%
\usepackage{makeidx}
\usepackage{amssymb}
\usepackage{dcolumn}
\usepackage{graphicx}
\usepackage{acronym}
\usepackage{amsmath}
\usepackage{amsfonts}%
\ifx\pdfoutput\relax\let\pdfoutput=\undefined\fi
\newcount\msipdfoutput
\ifx\pdfoutput\undefined\else
\ifcase\pdfoutput\else
\ifx\paperwidth\undefined\else
\ifdim\paperheight=0pt\relax\else\pdfpageheight\paperheight\fi
\ifdim\paperwidth=0pt\relax\else\pdfpagewidth\paperwidth\fi

\begin{document}
	
\title{Energy-momentum and Ricci tensors in a rotating space}
\author{B. V. Gisin }

\affiliation{E-mail: borisg2011@bezeqint.net}
\date{\today }
	
\begin{abstract}
Assuming that the Universe is an object of point rotation at a relic frequency, solutions of the Einstein equation are considered inside matter.

\textbf{Keywords}: Transformation for point rotating frames, Rotating universe, Cylindrical solutions in the general theory of relativity.  
\end{abstract}.
	
\maketitle
	
\section{Introduction}

Recently, the concept of the Universe as an object of point rotation was presented. In this concept, space rotates like a circularly polarized electromagnetic field at a frequency, called for  certainty, the relic frequency \cite{pr}. The frequency squared is proportional to the cosmological constant. The huge rotation period associated with the relic frequency does not affect ordinary life. The double wavelength corresponding to the frequency determines the size of the Universe.
  
A distinctive feature of the concept is using in the interval of the general theory relativity three-dimensional quadratic form, invariant under the three-dimensional transformation for rotating reference systems, which is a generalization of the Lorentz transformation.

In  \cite{pr}, cylindrical solutions of the Einstein equation with a cosmological term in free space are considered and it is shown that there are no solutions depending only on the cylindrical radius, but there is an exact solution depending on other coordinates, provided that the radius is an invariable parameter. 

However, the solution does not exclude the existence of singularities of the space in the form of four-dimensional astronomical objects, in time interval significantly smaller than the period of rotation and with the dimensions which are significantly smaller than the wavelength corresponding to the relic frequency. The space outside the objects can be considered as a flat Minkowski space.

The dualism inherent in this construction is similar to the principle "wave-particle" in quantum theory. On the one hand, the Universe has a size, as mentioned above, on the other hand, the Universe is infinite, because the axis of rotation can be selected at each point

In the paper we consider the space with the energy-momentum tensor. The  tensor is constructed in the standard way. 

The key point in simplifying the Einstein equation inside matter is the assumption that the cylindrical radius is an invariable parameter. This parameter is involved in the calculations, but drops out of the final results, remaining uncertain. Since the radius does not change, the radial component of the 4-velocity is considered to be zero. 

This paper is an extension of the set papers devoted to study of the three-dimensional transformation for rotating coordinate system as a generalization of the Lorentz transformation \cite{pr}. 

\subsection{The three-dimensional transformation}

3D transformation can be written as follows
\begin{align}
d\tilde{\varphi}  &  =d\varphi+\omega dz-\omega dt,\label{tphi}\\
d\tilde{z}  &  =-\frac{r^{2}\omega}{\varkappa}d\varphi+\kappa_{22}dz+ \kappa_{23}dt,\label{tz}\\
d\tilde{t}  &  =-\frac{r^{2}\omega}{\varkappa}d\varphi+\kappa_{32}%
dz+\kappa_{33}dt. \label{tt}%
\end{align} 
where $\varphi, z, t$ is the cylindrical angle, coordinate along the axis of rotation, time in the normalized system, where the speed of light is equal 1. $r$ is invariable parameter, $\varkappa$ is the constant, $ \kappa_ {kl} $ is defined as follows
\begin{align}
\kappa_{22}  &  =\frac{1}{2\varkappa}(1+\varkappa^{2}-\omega^{2}r^{2}), \label{C22} \\
\kappa_{23}  &  =\frac{1}{2\varkappa}(1-\varkappa^{2}+\omega^{2}r^{2}), \label{C23} \\
\kappa_{32}  &  =\frac{1}{2\varkappa}(1-\varkappa^{2}-\omega^{2}r^{2}), \label{C32}\\
\kappa_{33}  &  =\frac{1}{2\varkappa}(1+\varkappa^{2}+\omega^{2}r^{2}). \label{C33}%
\end{align}
Determinant of transformation is equal to 1. For $\omega=0$, (\ref{tphi})-(\ref{tt}) turns into the Lorentz transformation.

The quadratic differential form 
\begin{equation} 
r^2d\varphi^2+dz^2-dt^2, \label{dE}
\end{equation}
is invariant with respect to the transformation, therefore the interval of the flat Minkowski space also has this property. Consequently, the flat Minkowski space can be an object of point rotation with the relic frequency.

\section{The Einstein equation}

Within matter, the Einstein tensor, together with the cosmological constant, must be equated to the energy-momentum tensor
\begin{align}
R_{\mu\nu}-\frac{1}{2}Rg_{\mu\nu}+g_{\mu\nu }\Lambda=T_{\mu\nu }.  \label{EL}   
\end{align}

The cosmological constant $\Lambda$ is extremely small, of the order of $10^{-52}m^{-2}$ \cite{Tan}. For simplicity, we omit the gravitational constant. This is always possible with normalization.

We use the interval containing the quadratic form
\begin{align}
& ds^2=gdr^2+f(r^2d\varphi^2+dz^2-dt^2), \label{cgf} 
\end{align}

Covariant and contra-variant components of the metric tensor in cylindrical coordinates are	
\begin{align}
&  g_{\mu\nu}= \begin{vmatrix}
g         &  0      & 0 & 0 \\  
0 & r^2f  & 0 & 0 \\
0 & 0  & f & 0 \\
0 & 0  & 0 & -f \\
\end{vmatrix}, \quad
g^{\mu\nu}=\begin{vmatrix}
1/g         &  0      & 0 & 0 \\  
0 & 1/r^2f  & 0 & 0 \\
0 & 0  & 1/f & 0 \\
0 & 0  & 0 & -1/f \\
\end{vmatrix}, \nonumber
\end{align}
where $g,f$ are some functions of coordinates $\varphi, z, t$ and all off-diagonal components of this tensor are zero.

Usually for the energy-momentum tensor the form
\begin{align}
 T_{\mu\nu}= (\rho + p)u_{\mu}u_{\nu}-pg_{\mu\nu}, \label{Tr}  
\end{align}
is used, where $\rho $ is the matter density, $ p $ is the pressure, $u_{\mu}=g_{\mu\nu}dx^{\nu}/ds$ is the 4-vector velocity, satisfying equality
\begin{align}
g^{\mu\nu}u_{\mu}u_{\nu}=g_{\mu\nu}u^{\mu}u^{\nu}=1. \label{1ds} 
\end{align}

With help of (\ref{Tr}) we get another form of the Einstein equation
\begin{align}
& R_{\mu\nu} =(\rho + p)u_{\mu}u_{\nu}+g_{\mu\nu}\wp,\quad \mu=\nu, \label{Rmm} \\
& R_{\mu\nu} =(\rho + p)u_{\mu}u_{\nu}, \quad \mu \neq \nu, \label{Rmn} \\ 
& \wp=\Lambda-\frac{1}{2}(\rho-p). \label{wp}
\end{align}

\subsection{Note on mass density}

As is known, the Bethe-Weizker semi-empirical mass formula allows approximating the mass of atomic nuclei \cite{W}.
\begin{align}
m = Zm_p + Nm_n- a\frac{(A - 2Z)^2}{A}+ E_b(A,Z), \label{m}
\end{align}
where $Z$ and $N$ is the number of protons and neutrons, $A = Z + N$ is the total number of nucleons, $m_p$ and $m_n$ are the rest mass of a proton and a neutron, respectively, $E_b$ the binding energy of the nucleus, $a$ and $E_b$ are small and determined empirically.

Eq. (\ref{m}) can be changed to following form
\begin{align}
m = Zm_{pe} + Nm_{ne} - 4a\frac{Z^2}{A}+E_b, \label{me}
\end{align}
where $m_{pe}$ and $m_{ne}$ is the effective mass of the proton and neutron in the nucleus.

Neglecting the small term $ E_b $ and dividing the equation by $ A $, we obtain approximately that the total mass density is equal to the sum of the mass density of protons and neutrons minus the square of the charge density of protons. With a decrease in volume, this density increases faster and can dominate.

Emphasize that this semi-empirical formula has been experimentally verified for 100 years. Therefore, when constructing the energy-momentum tensor in the presence of an electromagnetic field, this additional negative term must be taken into account. As Einstein said "Our problem now is introduce a tensor $T_{\mu\nu}$, of the second rank, whose structure we do
know but provisionally ..." \cite{E}. 

This additional term can be associated with charged dark matter, discussed in the astrophysical community for a long time, and can be used for description of four-dimensional astronomical objects as singularities of the point rotating space. 

\subsection{Solutions in cylindrical coordinates}   

The concept of point rotation assumes that the cylindrical radius is an invariable parameter. This means that the corresponding component of the 4-vector $ u ^ r $ is zero by definition.

It follows from the equation (\ref{Rmn}) that all off-diagonal components of the Ricci tensor with index $ r $ are zero. Because of this
\begin{align}
g=fC(\varphi), \label{fg}
\end{align}
where $f$ is a function only of $z, t$ and $C$ is a function only of $\varphi$.
Using this relations and the expressions for components of the Ricci tensor, easily to shown that
\begin{align}
R_{\varphi z} \equiv 0, \quad R_{\varphi t} \equiv 0. \label{Rvvzt} 
\end{align}

It can be the shown that the components $fR_{rr}/g$ and $R_{\varphi\varphi}/r^2$ expressed in terms of the metrical tensor coincide. Accordingly to (\ref{Rmm}) the difference of the component is equal 
$(\rho +  p)u^2_{\varphi}$, since $u_{r}=0$. The case $(\rho+p)=0$ reduced the energy-momentum tensor to the form 
\begin{align}
T_{\mu\nu}=-pg_{\mu\nu}=\rho g_{\mu\nu}, \label{ghr}
\end{align}
where $p, \rho$ must be constant, because the covariant derivative of the tensor must be zero.

In this case the problem is equivalent to solutions for the free space, where 
the role of cosmological constant plays $\Lambda+p=\Lambda-\rho$. This case is detailed below.

We assume that $u_{\varphi}=0$. It means that only components $u_{z}, u_{t}$ are not equal zero.

The term describing dependence $R_{rr}$ on derivatives with respect to $\varphi$ is
\begin{align} 
& \delta_{\varphi}=-\frac{C_{,\varphi\varphi}}{2r^2C}+ \frac{C^2_{,\varphi}}{4r^2C^2}.
\end{align} 
Combinations of the Ricci tensor components can be written so that the right site of the equation (\ref{Rmm}) contains the term $(\rho + p)u^2_{\mu}$
\begin{align}  
& R_{zz}-\frac{f}{g}R_{rr}:-\delta_{\varphi}-(\frac{2}{\sqrt{f}})_{,zz}=
(\rho + p)u^2_{z}, \nonumber \\
& R_{tt}+\frac{f}{g}R_{rr}:+\delta_{\varphi}-(\frac{2}{\sqrt{f}})_{,tt}=
(\rho + p)u^2_{z}. \nonumber \\
& R_{zt}:-(\frac{2}{\sqrt{f}})_{,zz}=(\rho + p)u_{z}u_{t}. \nonumber 
\end{align} 
Using these expressions, a differential equation for $f$ can be written as follows
\begin{align}
[(\frac{4}{\sqrt{f}})_{,zt}]^2=[\delta_{\varphi}+(\frac{2}{\sqrt{f}})_{,zz}] [-\delta_{\varphi}+(\frac{2}{\sqrt{f}})_{,tt}]. \label{zt}
\end{align}
From this equation immediately follows that  
\begin{align}
\delta_{\varphi}=0, \quad \sqrt{C(\varphi)}=a\varphi+b, \label{ab}
\end{align}
where $a, b$ are constants, $f$ is an arbitrary function of $\eta$,  $\eta=kz+\nu t+c_0$ and $k, \nu, c_0$ are constants.  

Expressions for diagonal components $fR_{rr}/g$ and $R_{\varphi\varphi}/r^2$ coincide and from (\ref{Rmm}) we get the first differential equation for $f$ 
\begin{align} 
-\frac{f_{,zz}}{2f}+\frac{g_{,tt}}{2g}=f\wp, \quad \text{or} \quad
(\nu^2-k^2)\frac{f_{,\eta\eta}}{2f}=f\wp. \label{zt1}
\end{align}
The contravariant component $u^{z}, u^{t}$ are connected by the equality
\begin{align}
f\frac{dz^2}{ds^2}-f\frac{dt^2}{ds^2} \equiv f(u^{z})^2-f(u^{t})^2=1. \label{zt3}
\end{align}
Consider three equations for the $f$ as a function $\eta$ originated from $R_{zz}$, $R_{tt}$ and $R_{zt}$. The equations can be written in the form 
\begin{align}  
& k^2(-\frac{f_{,\eta\eta}}{f}+\frac{3f^2_{,\eta}}{2f^2})=
(\rho + p)u_{z}u_{z}, \label{Rzz} \\
& \nu^2(-\frac{f_{,\eta\eta}}{f}+\frac{3f^2_{,\eta}}{2f^2})=
(\rho + p)u_{t}u_{t}, \label{Rtt1} \\
& k\nu(-\frac{f_{,\eta\eta}}{f}+\frac{3f^2_{,\eta}}{2f^2})=
(\rho + p)u_{z}u_{t}. \label{Rzt1} 
\end{align}
From the equations  we define components of 4-velocity
\begin{align}  
& (u^{t})^2=\frac{\nu^2}{f(\nu^2-k^2)}, \quad (u^{z})^2=\frac{k^2}{f(\nu^2-k^2)}, \nonumber \\
& u^{t}u^{z}=\frac{k\nu}{f(\nu^2-k^2)}. \nonumber
\end{align}
The constants $\nu, k$ must satisfy the condition $\nu^2>k^2$.

With this definitions above, three differential equations turn out into one
\begin{align}  
(\nu^2-k^2)(-f_{,\eta\eta}+\frac{3f^2_{,\eta}}{2f})=(\rho + p). \label{eta}
\end{align}

To determine the three functions $ f, \rho, p $ we have two equations. The next section shows that they can be compatible and define conditions for that.

For more information about the conditions, consider the equation of motion
 $T^{\mu\nu}_{;\nu}=0$.

\subsection{The equation of motion}

For the case $u_{z}\neq 0, u_{t}\neq 0$ there are only three nonzero components of the energy-momentum tensor $T^{zz}, T^{zt}, T^{tt}$. The three components form two equations of motions describing the conservation of momentum and energy at $\mu=z$ and $\mu=t$

Substituting expressions of the energy-momentum tensor into the equations, we obtain two necessary conditions for the existence of solutions 
\begin{align}
& [\rho(\nu^2+k^2)+2pk^2]_{,\eta}+\frac{f_{,\eta}}{2f}[(\rho + p)(3\nu^2+k^2)]=0, \nonumber \\ 
& [\rho(\nu^2+k^2)+2p\nu^2]_{,\eta}+
\frac{f_{,\eta}}{2f}(\rho + p)(\nu^2+3k^2)=0. \nonumber 
\end{align}

Using the inequality $\nu^2 > k^2$, take the difference and sum of the equations
\begin{align}
\frac{f_{,\eta}}{f}(\rho + p)-2p_{,\eta}=0, \;\;\;
 \frac{f_{,\eta}}{f}(\rho + p)+(\rho+p)_{,\eta}=0. \nonumber 
\end{align}
Two condition follow from this
\begin{align}
\rho+3p=\alpha, \quad (\rho+p)f=\beta, \label{AB} 
\end{align}
where $\alpha, \beta$ are constants.

For compatibility of equations (\ref{zt3}) and (\ref{eta}), as in the previous case, the condition must be satisfied
\begin{align}
\rho+p=0,  \label{AB1} 
\end{align}
must be satisfied, where $p, \rho=-p$ and $\wp=\Lambda+p$ are also constants.

The solution for both the equations is
\begin{align}
& f=\frac{1}{(\gamma\eta+\eta_1)^2}, \quad
 \gamma^2=\frac{(\Lambda+p)}{3(\nu^2-k^2)}. \label{f} 
\end{align}
The constant $\eta_1$ is define the origin of the coordinates and may be neglected. 

\section{Conclusion}

The assumption that the cylindrical radius is invariable and, accordingly, the radial component of the 4-velocity is zero, uniquely determines exact solutions of the Einstein equation. These solutions completely equivalent to the case of free space, where the difference between the cosmological constant and the mass density plays the role of this constant. Mass density and pressure should have opposite signs and be constant.

\end{document}